\documentclass[12pt,preprint]{aastex}
\usepackage{emulateapj5,apjfonts}
\usepackage{graphics}
\usepackage{amssymb}
\usepackage{bbm}
\usepackage{amsmath,textcomp,array}
\usepackage{onecolfloat}
\usepackage{bm}
\lefthead{Kwan et al.}
\righthead{Concentration Mass Relation Predictions for wCDM Universes}

\begin{document}


\def\head{

\title{Cosmic Emulation: The Concentration-Mass Relation for $w$CDM Universes}
\author{Juliana Kwan\altaffilmark{1}, Suman Bhattacharya\altaffilmark{1,2}, Katrin Heitmann\altaffilmark{1,2,3,4},
and Salman Habib\altaffilmark{1,2,3,4}}

\affil{$^1$ High Energy Physics Division, Argonne National Laboratory, Lemont, IL 60439}
\affil{$^2$ Kavli Institute for Cosmological Physics, The University of Chicago, Chicago, IL 60637 }
\affil{$^3$ Mathematics and Computer Science Division, Argonne National Laboratory, Lemont, IL 60439}
\affil{$^4$ Computation Institute, The University of Chicago, Chicago, IL 60637}

\date{today}

\begin{abstract}
  The concentration-mass relation for dark matter-dominated halos is
  one of the essential results expected from a theory of structure
  formation. We present a simple prediction scheme, a {\em cosmic
    emulator}, for the $c-M$ relation as a function of cosmological
  parameters for $w$CDM models. The emulator is constructed from 37
  individual models, with three nested $N$-body gravity-only
  simulations carried out for each model.  The mass range covered by
  the emulator is ${2\cdot 10^{12}{\rm M}_{\odot}<{\rm M}<10^{15}{\rm
      M}_{\odot}}$ with a corresponding redshift range of $z=0 -
  1$. Over this range of mass and redshift, as well as the variation
  of cosmological parameters studied, the mean halo concentration
  varies from $c\sim 2$ to $c\sim 8$. The distribution of the
  concentration at fixed mass is Gaussian with a standard deviation of
  one-third of the mean value, almost independent of cosmology, mass,
  and redshift over the ranges probed by the simulations. We compare
  results from the emulator with previously derived heuristic analytic
  fits for the $c-M$ relation, finding that they underestimate the
  halo concentration at high masses. Using the emulator to investigate
  the cosmology dependence of the $c-M$ relation over the currently
  allowable range of values, we find -- not surprisingly -- that
  $\sigma_8$ and $\omega_m$ influence it considerably, but also that
  the dark energy equation of state parameter $w$ has a substantial
  effect. In general, the concentration of lower-mass halos is more
  sensitive to changes in cosmological parameters as compared to
  cluster mass halos.

  The $c-M$ emulator is publicly available from {\tt
    http://www.hep.anl.gov/cosmology/CosmicEmu}.
\end{abstract}

\keywords{methods: statistical ---
          cosmology: large-scale structure of the universe}}

\twocolumn[\head]

\section{Introduction}

Over the last three decades, cosmology has made tremendous progress,
culminating in the so-called ``Standard Model of Cosmology''. The two
main components of the Standard Model are a mysterious dark energy,
leading to a late-time accelerated expansion, and a dark matter
component making up roughly 25\% of the total matter-energy budget of
the Universe. The evolution of structure in the Universe from the
earliest accessible times to today is successfully described by a
theory based on the gravitational instability -- the distribution of
galaxies in the Universe, for example, is remarkably well reproduced
by this paradigm. Clusters and groups of galaxies are major building
blocks of the large-scale structure and measurements of their
abundance provide a powerful cosmological probe. Large, gravity-only
$N$-body simulations have been remarkably successful in providing a
consistent picture of the formation of the large-scale structure from
the very early, small Gaussian density fluctuations to the halos,
voids, and filaments we observe today.

A surprising discovery from these simulations~\citep{nfw1,nfw2} was
that the dark matter-dominated halos -- over a wide mass range typical
of dwarf galaxies to massive clusters -- share a basically universal
density profile. In detail, it was shown that the spherically averaged
density profile of relaxed halos formed in simulations can be
described by what is now commonly known as the NFW
(Navarro-Frenk-White) profile. The NFW profile is described by two
parameters, the normalization and the characteristic scale radius of
the halo or equivalently its (dimensionless) concentration.

Aside from the distribution of halo masses, halo profiles are also of
considerable interest. The profiles can be measured directly for
individual massive halos by a variety of observational methods, or
inferred indirectly for less massive halos using statistical lensing
probes. Halo profiles are also a key input in halo occupation
distribution (HOD) modeling of the distribution of galaxies. NFW
profiles (or minor variants thereof) are consistent with current
observations~\citep{bhattacharya11_b} and, as the observations continue
to improve, a corresponding improvement in theoretical predictions for
(NFW) halo concentrations as a function of cosmological parameters is
needed. As scatter in the $c-M$ relation is considerable, in
principle, this would encompass knowing the actual distribution of
halo concentrations as a function of halo mass.

Quantitative predictions for the $c-M$ relation from a first
principles analytic approach are difficult to obtain, due to the
highly nonlinear dynamics involved in the formation of halos. Accurate
predictions can only be obtained from computationally expensive,
high-resolution simulations. These simulations need to cover large
volumes in order to yield good statistics, especially in the cluster
mass regime, as well as high force resolution to reliably resolve the
halo profiles. In recent years, the focus has therefore been on
generating predictions for one cosmology around the best-fit WMAP
(Wilkinson Microwave Anisotropy Probe) results of that time~(e.g.,
\citealt{duffy08}, \citealt{bhattacharya11_b},
\citealt{prada12}). The fitting functions so generated cannot be
extended beyond the cosmological model they have been tuned
for. Heuristic models that aim to extend this reach, e.g. those by
\cite{bullock} and \cite{eke01} and improvements thereof
(\citealt{maccio08}) do not lead to the desired accuracy, as discussed
in \cite{duffy08}. Other discussions of this issue can be found in
~\cite{gao07}, \cite{hayashi07}, and \cite{zhao09}.

In order to overcome the many shortcomings of fitting functions as a
general approach in cosmology, we have recently developed the ``Cosmic
Calibration Framework'' (CCF) to provide accurate prediction schemes
for cosmological observables~\citep{HHHN,HHHNW}. The aim of the CCF is
to build codes that act as very fast -- basically instantaneous --
prediction tools for large scale structure observables such as the
nonlinear power spectrum~\citep{coyote1,coyote2,coyote3}, mass
functions for different halo definitions, or the concentration-mass
relation -- as discussed here. Predicting these observables requires
running a number of high-performance simulations to reliably resolve
the nonlinear regime of structure formation. The CCF provides a
powerful way to build precision prediction tools from a limited
number of computationally expensive simulations.

At the heart of the CCF lies a sophisticated sampling scheme that
provides an optimal sampling strategy for the cosmological models to
be simulated (we use orthogonal array-based Latin hypercube as well as
symmetric Latin hypercube designs; an introduction to the general
sampling strategy is provided in~\citealt{santner03}), an optimal
representation to translate the measurements from the simulations into
functions that can be easily interpolated (a principal component basis
turns out to be an efficient representation), and finally a very
accurate interpolation scheme (our choice here is Gaussian process
modeling).  

The CCF was first introduced in~\cite{HHHN} and a more detailed
description and examples are provided in~\cite{HHHNW}. In a series of
three papers (Coyote Universe I-III) we developed an emulator for the
matter power spectrum for a five dimensional parameter space covering
$\theta=\{\omega_b, \omega_m, n_s, w, \sigma_8\}$. This emulator
provides predictions for the power spectrum for $w$CDM cosmologies out
to $k\sim 1$~Mpc$^{-1}$ at the 1\% accuracy level for a redshift range
of $0\le z\le 1$. In~\cite{SKHHHN} the work was extended to derive an
approximate statistical model for the sample variance distribution of
the nonlinear matter power spectrum. \cite{eifler} used the emulator
to generate a weak lensing prediction code to calculate various
second-order cosmic shear statistics, e.g., shear power spectrum,
shear-shear correlation function, ring statistics and Complete
Orthogonal Set of EB-mode Integrals (COSEBIs).

The focus of this paper is the development of an emulator for the
concentration-mass relation for $w$CDM cosmologies. We use the same
base set of simulations as in \cite{coyote3}, consisting of 37
cosmological models and a single 1300~Mpc volume, high-resolution
simulation for each model. This simulation set is augmented here with
a set of new, higher resolution simulations. These simulations cover
smaller volumes (a 360~Mpc and a 180~Mpc simulation for each model) to
obtain good statistics over a large range of halo masses. For each
model we measure the best-fit concentration-mass ($c-M$) relation,
assuming a simple power law form. The fits lay the foundation for
building the emulator that provides predictions for the $c-M$ relation
within the $w$CDM parameter space covered by the original simulations.
In redshift, the emulator covers the range between $z=0$ and $z=1$. We
provide a fast code that delivers the mean $c-M$ relations for $w$CDM
cosmologies to good accuracy\footnote{{\tt
    http://www.hep.anl.gov/cosmology/CosmicEmu}}.

As is well-known, the $c-M$ relation has considerable scatter and, in
principle, it is not obvious that this scatter should have a simple
form, and what its cosmological dependence might be. However, as
discussed in~\cite{bhattacharya11_b}, the scatter has a simple
Gaussian form in $w$CDM models, and moreover, even though the mean
$c-M$ relation is clearly cosmology-dependent, as is the associated
concentration variance, $\sigma_c^2(M)$, the ratio of $\sigma_c(M)$ to
the mean concentration is close to $1/3$, independent of cosmology,
mass, or redshift. This means that once an emulator for the $c-M$
relation is in hand, the concentration standard deviation is given
automatically by a simple relation.

The paper is organized as follows. After a brief outline of the halo
concentration measurements from the simulations, we describe the
cosmological model space and the simulation suite used to build the
emulator. In Section~\ref{sec:models} we also discuss the generation
of the smooth prediction for the concentration-mass relation for each
model that underlies the interpolation scheme for building the
emulator. We give a brief description on how to build the emulator in
Section~\ref{sec:emu} and show some examples from the working emulator
and test results verifying its accuracy. We also compare our results
to currently used fitting formulae and investigate the cosmology
dependence of the $c-M$ relation in some detail.  Finally, we provide
a conclusion and outlook in Section~\ref{sec:conc}.

\section{Concentration-Mass Relation} \label{sec:cm}

We study the concentration-mass relation in the regime of bright
galaxies to clusters of galaxies, spanning halo mass ranges between
$2\cdot 10^{12}$M$_\odot$ to $10^{15}$M$_\odot$, while varying $w$CDM
cosmological parameters. A detailed description on how to measure halo
concentrations from simulations and a discussion of possible
systematics is given in~\cite{bhattacharya11_b}. We follow the same
approach in this paper and give here a brief summary of the main steps
in measuring the $c-M$ relation in our simulations.

As a first step, we identify halos using a fast parallel
friends-of-friends (FOF) finder \citep{woodring11} with linking length
$b=0.2$. Once a halo is found, we define its center via a density
maximum criteria -- the location of the particle with the maximum
number of neighbors. This definition of the halo center is very close
to that given by the halo's potential minimum. Given a halo center, we
grow spheres around it and compute the mass in radial bins. Note that
even though an FOF finder is used, the actual halo mass is defined by
a spherical overdensity method, consistent with what is done in
observations. (For discussions on halo mass, see, e.g.,
\citealt{white01}, \citealt{lukic09}, and \citealt{more11}). The NFW
form for the spherically averaged halo profile is a function of two
parameters, one of which is constrained by the halo mass. Here we fit
the mass profile using both total halo mass and concentration as
free variables. Although the mass could be measured independently of
the concentration, the joint analysis is potentially less sensitive to
fitting bias.

We write the NFW profile as
\begin{equation}
\rho(r)= \frac{\delta\rho_{\rm{crit}}}{(r/r_s)(1+r/r_s)^2},
\label{eq:nfw}
\end{equation}
where $\delta$ is a characteristic dimensionless density, and $r_s$ is
the scale radius of the NFW profile. The concentration of a halo is
defined as $c_{\Delta}=r_{\Delta}/r_s$, where $\Delta$ is the
overdensity with respect to the {\it critical density} of the
Universe, $\rho_{\rm{crit}}=3H^2/8\pi G$, and $r_{\Delta}$ is the
radius at which the enclosed mass, $M_{\Delta}$, equals the volume of
the sphere times the density $\Delta \rho_{\rm{crit}}$. We compute
concentrations corresponding to $\Delta=200$, corresponding in turn to
$c_{200}= R_{200}/r_s$.

\begin{table*}
\begin{center} 
\caption{The parameters for the 37+1 models which define the sample
  space. See text for further details. \label{tab:basic}}
\vspace{-0.3cm}
\begin{tabular}{ccccccc|ccccccc}
\# & $\omega_m$ & $\omega_b$ & $n_s$ & $-w$ & $\sigma_8$ & $h$ &
\# & $\omega_m$ & $\omega_b$ & $n_s$ & $-w$ & $\sigma_8$ & $h$ \\ \hline 
 M000 & 0.1296 & 0.0224 & 0.9700 & 1.000 & 0.8000 & 0.7200  &
M019 & 0.1279 & 0.0232 & 0.8629 & 1.184 & 0.6159 & 0.8120  \\ 
 M001 & 0.1539 & 0.0231 & 0.9468 & 0.816 & 0.8161 & 0.5977 & 
M020 & 0.1290 & 0.0220 & 1.0242 & 0.797 & 0.7972 & 0.6442  \\ 
 M002 & 0.1460 & 0.0227 & 0.8952 & 0.758 & 0.8548 & 0.5970 & 
M021 & 0.1335 & 0.0221 & 1.0371 & 1.165 & 0.6563 & 0.7601  \\ 
 M003 & 0.1324 & 0.0235 & 0.9984 & 0.874 & 0.8484 & 0.6763 & 
M022 & 0.1505 & 0.0225 & 1.0500 & 1.107 & 0.7678 & 0.6736  \\ 
 M004 & 0.1381 & 0.0227 & 0.9339 & 1.087 & 0.7000 & 0.7204 & 
M023 & 0.1211 & 0.0220 & 0.9016 & 1.261 & 0.6664 & 0.8694 \\ 
M005 & 0.1358 & 0.0216 & 0.9726 & 1.242 & 0.8226 & 0.7669 & 
M024 & 0.1302 & 0.0226 & 0.9532 & 1.300 & 0.6644 & 0.8380 \\ 
M006 & 0.1516 & 0.0229 & 0.9145 & 1.223 & 0.6705 & 0.7040 & 
M025 & 0.1494 & 0.0217 & 1.0113 & 0.719 & 0.7398 & 0.5724 \\ 
 M007 & 0.1268 & 0.0223 & 0.9210 & 0.700 & 0.7474 & 0.6189 &
M026 & 0.1347 & 0.0232 & 0.9081 & 0.952 & 0.7995 & 0.6931  \\ 
 M008 & 0.1448 & 0.0223 & 0.9855 & 1.203 & 0.8090 & 0.7218 & 
M027 & 0.1369 & 0.0224 & 0.8500 & 0.836 & 0.7111 & 0.6387  \\ 
 M009 & 0.1392 & 0.0234 & 0.9790 & 0.739 & 0.6692 & 0.6127 & 
M028 & 0.1527 & 0.0222 & 0.8694 & 0.932 & 0.8068 & 0.6189  \\ 
M010 & 0.1403 & 0.0218 & 0.8565 & 0.990 & 0.7556 & 0.6695 & 
M029 & 0.1256 & 0.0228 & 1.0435 & 0.913 & 0.7087 & 0.7067  \\ 
M011 & 0.1437 & 0.0234 & 0.8823 & 1.126 & 0.7276 & 0.7177 & 
M030 & 0.1234 & 0.0230 & 0.8758 & 0.777 & 0.6739 & 0.6626  \\ 
M012 & 0.1223 & 0.0225 & 1.0048 & 0.971 & 0.6271 & 0.7396 & 
M031 & 0.1550 & 0.0219 & 0.9919 & 1.068 & 0.7041 & 0.6394  \\ 
M013 & 0.1482 & 0.0221 & 0.9597 & 0.855 & 0.6508 & 0.6107 & 
M032 & 0.1200 & 0.0229 & 0.9661 & 1.048 & 0.7556 & 0.7901  \\
M014 & 0.1471 & 0.0233 & 1.0306 & 1.010 & 0.7075 & 0.6688 & 
M033 & 0.1399 & 0.0225 & 1.0407 & 1.147 & 0.8645 & 0.7286  \\ 
M015 & 0.1415 & 0.0230 & 1.0177 & 1.281 & 0.7692 & 0.7737 & 
M034 & 0.1497 & 0.0227 & 0.9239 & 1.000 & 0.8734 & 0.6510  \\  
M016 & 0.1245 & 0.0218 & 0.9403 & 1.145 & 0.7437 & 0.7929 & 
M035 & 0.1485 & 0.0221 & 0.9604 & 0.853 & 0.8822 & 0.6100  \\ 
M017 & 0.1426 & 0.0215 & 0.9274 & 0.893 & 0.6865 & 0.6305 & 
M036 & 0.1216 & 0.0233 & 0.9387 & 0.706 & 0.8911 & 0.6421  \\
M018 & 0.1313 & 0.0216 & 0.8887 & 1.029 & 0.6440 & 0.7136 & 
M037 & 0.1495 & 0.0228 & 1.0233 & 1.294 & 0.9000 & 0.7313 
\end{tabular}
\end{center}
\end{table*}

The mass enclosed within a radius $r$ for an NFW halo is given by 
\begin{equation}
M(<r)= m(c_\Delta r/R_{200})/m(c_{200})M_{200},
\label{eq:nfwmass}
\end{equation}
where $m(y)=\ln(1+y)-y/(1+y)$. 
The mass in a radial bin is then
\begin{equation}
M_i=M(<r_i)-M(<r_{i-1}).
\label{eq:nfwmassbin}
\end{equation}
We then fit Eq.~\ref{eq:nfwmassbin} to the mass contained in the radial
bins of each halo, by minimizing the associated value of $\chi^2$ as
\begin{equation}
\chi^2= \sum_i \frac{(M_i^{sim}-M_i)^2}{(M_i^{sim})^2/n_i}, 
\label{eq:fit}
\end{equation}
where the sum is over the radial bins, $n_i$ is the number of
particles in a radial bin, $M_i^{sim}$ is the mass in bin $i$
calculated from the simulations and $M_i$ is the mass calculated
assuming the NFW profile. The advantage of fitting mass in radial bins
rather than the density is that the bin center does not have to be
specified. Note that we explicitly account for the finite number of
particles in a bin. This leads to a slightly larger error in the
profile fitting but minimizes any possible bias due to the finite
number of particles, especially near the halo center.

We fit for two parameters -- the normalization of the profile and the
concentration. Halo profiles are fitted in the radial range of
approximately $(0.1-1)R_{vir}$. This choice is motivated partly by the
observations of concentrations that typically exclude the central
region of clusters (e.g., observations by \citealt{oguri11}). More
significantly, however, this excludes the central core which is
sensitive to the effects of baryonic physics and numerical errors
arising from limitations in both mass and force
resolution. \cite{duffy10} have shown that, at $r<0.1R_{vir}$, cluster
halo profiles are potentially sensitive to the impact of baryons with
the profiles being affected at $r=0.05 R_{vir}$ by as much as a factor
of 2.

The $c-M$ relation is calculated  by weighing the individual
concentrations by the halo mass, 
\begin{equation}
c(M)= \frac{\sum_i c_i M_i}{\sum_i M_i},
\label{eq:mean_c}
\end{equation}
where the sum is over the number, $N_i$, of the halos in a mass
bin. The mass of the bin is given by 
\begin{equation}
M= \sum_i M_i/N_i.
\label{eq:m}
\end{equation}
The error on $c(M)$ is the mass-weighted error on the individual fits
plus the Poisson error due to the finite number of halos in an
individual bin added in quadrature,
\begin{equation}
\Delta c(M)= \sqrt {\left (\frac{\sum_i \Delta c_i M_i}{\sum_i
      M_i}\right )^2+ \frac{c^2(M)}{N_i}},
\label{eq:errmean}
\end{equation} 
where $\Delta c_i$ is the individual concentration error for each
halo. The first term dominates towards the lower mass end where the
individual halos have smaller number of particles and the second term
dominates towards the higher mass end, where there are fewer halos to
average over.

\section{Cosmological Models and Simulation Sets}
\label{sec:models}

\begin{figure*}[t]
\centerline{
 \includegraphics[width=7.5in]{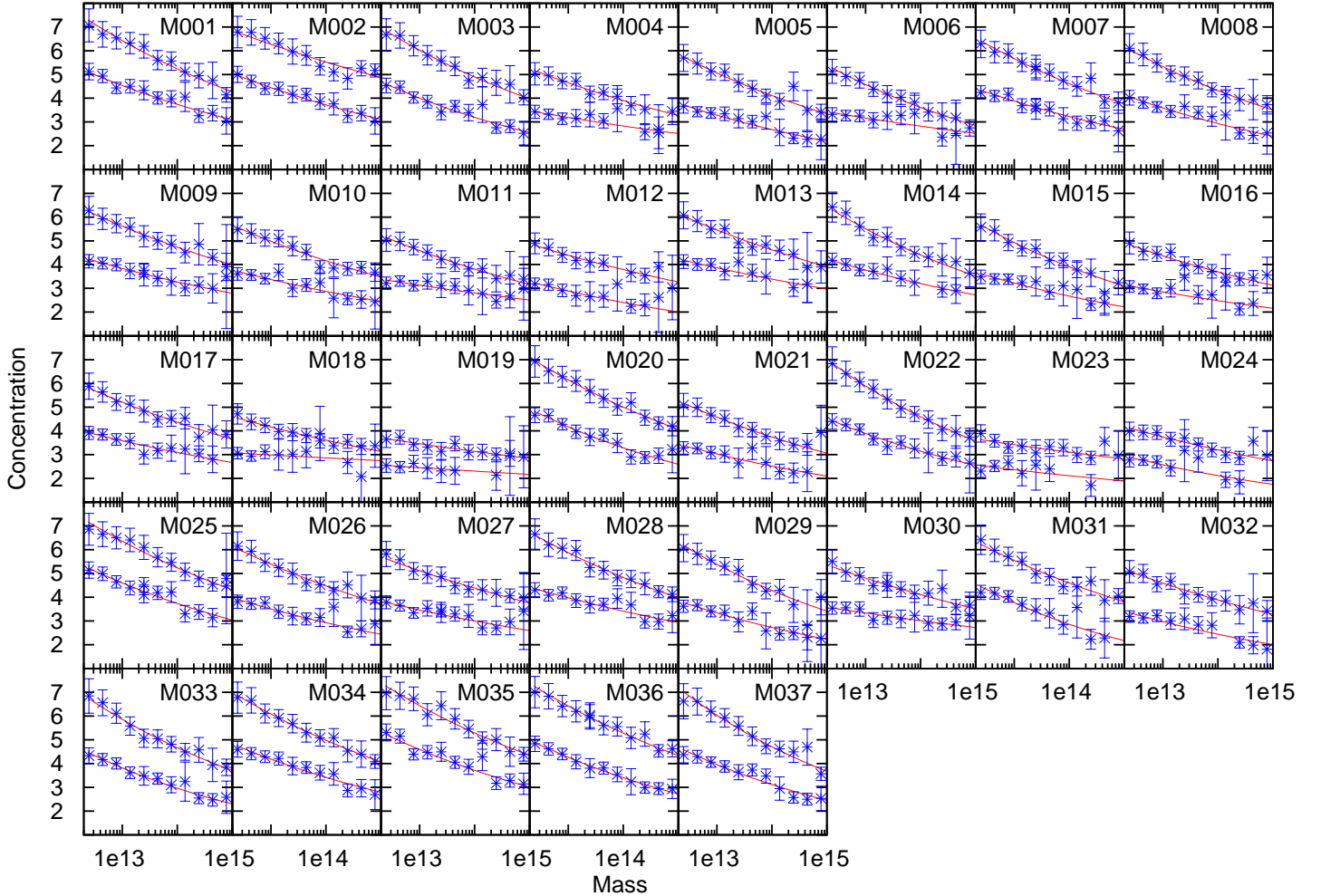}}
\caption{\label{cm-rel}Concentration-mass relations for 37 $w$CDM
  cosmologies. The blue points show the measurements from the three
  simulations per model while the red lines show the best-fit power
  law for each measurement. In each subplot we show the results for
  $z=0$ and $z=1$ (upper and lower curve respectively). We find the
  best-fit power law separately for both redshifts. Models with low
  values for $\sigma_8$ in general also exhibit lower $c-M$ relations
  (e.g. M012, M018, M019). The fits shown here are the foundation for
  building the emulator described in Section~\ref{sec:emu}. }
\end{figure*}

We now describe the cosmological model space covered by our prediction
scheme and the simulations used to construct it. The emulator is based
on 37 cosmological models spanning the class of $w$CDM cosmologies. We
allow for variations of the following five parameters:
\begin{equation}
\theta=\{\omega_b, \omega_m, n_s, w, \sigma_8\}.
\end{equation}
The 37 models are chosen to lie within the ranges:
\begin{equation}
\begin{array}{c}
0.0215 < \omega_b < 0.0235, \\
  0.120 < \omega_m < 0.155, \\
  0.85  < n_s      < 1.05, \\
  -1.30 < w        <-0.70, \\
  0.616  < \sigma_8 < 0.9,
\end{array}
\label{priors}
\end{equation}
which are picked based on current constraints from CMB
measurements~\citep{wmap7}. Following the approach in~\cite{coyote3}
we lock the value of the Hubble parameter $h$ to the best-fit value
for each model, given the measurement of the distance to the surface
of last scattering.  The values for $h$ then range from $0.55 < h <
0.85$.  In addition to the 37 models, we run one $\Lambda$CDM model
(M000 in Table~\ref{tab:basic}) which is not used to build the
emulator. Instead we use this model as a control for testing the
accuracy of the emulator. All 37+1 models are specified in detail in
Table~\ref{tab:basic}.

The specific model selection process is described at length in
\cite{coyote2}. In summary, it is based on Symmetric Latin Hypercube
(SLH) sampling~\citep{slh}; this sampling strategy provides a scheme
that guarantees good coverage of the parameter hypercube. In our
specific case we choose an SLH design that has good space filling
properties in the case of two-dimensional projections in parameter
space. In other words, if any two parameters are displayed in a plane,
the plane will be well covered by simulation points. \cite{coyote2}
provide an extensive discussion regarding optimal design choices and
we refer the interested reader to that paper.

\begin{table}
\begin{center} 
  \caption{Box sizes, particle numbers, and mass resolution. \label{tab:nest}}
\begin{tabular}{ccccc}
Length $[$Mpc$]$  & $N_p^3$ & Force res.  [kpc] & $m_p$ $[$M$_\odot]$ \\
\hline
1300 & $1024^3$  & 50 & $5.7\cdot10^{11}\omega_m $  \\
365   & $512^3$    & 10 & $1.0\cdot 10^{11}\omega_m $ \\
180   & $512^3$    & 10 & $1.2\cdot 10^{10}\omega_m $ \\
\end{tabular}
\end{center}
\end{table}

The emulator developed here is valid between $0<z<1$ and covers a halo
mass range from $2\cdot 10^{12}$M$_\odot$ to $10^{15}$M$_\odot$.  We
use different box sizes to cover different mass ranges with sufficient
statistics.  A summary of the different simulation sizes is given in
Table~\ref{tab:nest}. All simulations were carried out with the TreePM
code {\sc GADGET-2}~\citep{gadget2}. In previous work
\citep{bhattacharya11_b}, we have shown that results from {\sc
  GADGET-2} simulations and those with HACC \citep{habib09,pope10}
produce completely consistent results. Results from a recent cluster
re-simulation campaign \citep{wu12} are also in good agreement with
those of \cite{bhattacharya11_b}.

One set of simulations is from the original Coyote Universe suite as
described in~\cite{coyote3}. This set of runs evolves 1024$^3$
particles in (1300~Mpc)$^3$ volumes. In addition, we run one
realization each per model with 512$^3$ particles with a 10~kpc force
resolution in a 365~Mpc box and a 180~Mpc box. A summary of the
simulation sets including force and mass resolution is given in
Table~\ref{tab:nest}. 

We combine the simulation results from the three boxes for each model
to obtain measurements spanning the desired mass range. While some
models (in particular those with high values of $\sigma_8$) have
clusters at even higher masses, the statistics beyond
10$^{15}$M$_\odot$ are insufficient and we exclude those
measurements. This is also done in order to avoid extrapolations for
models where no data points at high masses exist. In order to build an
emulator, for each model we have to provide a prediction for the $c-M$
relation for the same mass range. This ensures that we can provide a
consistent set of measurements for the final interpolation process
between different models. From the simulation results, we determine
for each of the 37+1 cosmologies the best-fit $c-M$ relation by simply
finding the best-fit power law for each model at two redshifts, $z=0$
and $z=1$. The results for the 37 models underlying the emulator are
shown in Fig.~\ref{cm-rel}. The blue points show the simulation
results while the red curves show the best-fit power law for each
model. The upper curves in each plot are obtained at redshift $z=0$
and the lower curves at $z=1$. The concentration values range between
$c\sim 2$ and $c\sim 8$. As expected, we find that models with low
values of $\sigma_8$ (e.g., M012, M018, M019 with $\sigma_8<0.65$)
have depressed $c-M$ relations.  We will return to the cosmology
dependence of the $c-M$ relation in Section~\ref{subsec:testing} after
constructing the emulator, which will allow us to carry out a
comprehensive sensitivity analysis. We reiterate that the fits shown
in Fig.~\ref{cm-rel} are the basis for building the emulator; this
procedure is discussed in the next section.
\begin{center}
\begin{figure}[t]
\includegraphics[width=\linewidth]{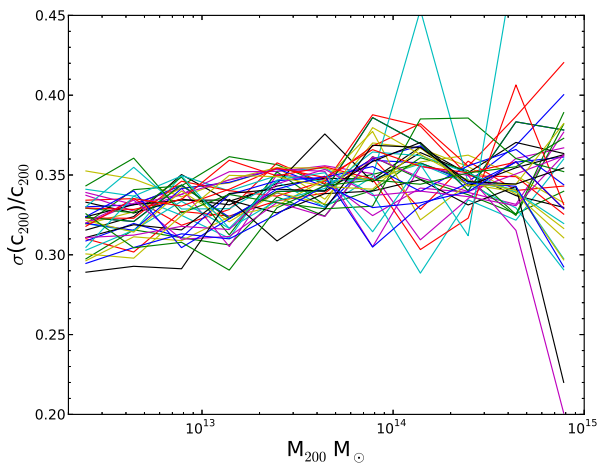}
\caption{Ratio of the standard deviation of the concentration to the
  mean concentration as a function of mass for all 37 cosmologies at
  $z=0$. While both the standard deviation and the mean concentration
  are functions of cosmology and redshift, their ratio is essentially
  invariant, and is approximately $1/3$. The distribution of the
  concentration around the mean is well-fit by a Gaussian
  distribution~\citep{bhattacharya11_b}.}
\label{fig:var}
\end{figure}
\end{center}

Finally we turn to a discussion of the intrinsic scatter in the $c-M$
relation. As mentioned earlier, the distribution of concentrations at
any given halo mass is Gaussian, and the ratio of $\sigma_c(M)$ to the
mean concentration is an approximate invariant for $w$CDM models, with
a value of $\sim 1/3$~\citep{bhattacharya11_b}, independent of redshift and
halo mass. This behavior is exhibited in Fig.~\ref{fig:var} where the
ratio is computed for all 37 cosmologies as a function of halo mass,
at $z=0$. Thus, given the $c-M$ relation from the emulator, the
standard deviation at each mass bin can be trivially estimated by
multiplying the returned concentration value by $1/3$.

\section{Emulator for the Concentration-Mass Relation}
\label{sec:emu}
\subsection{Building the emulator}\label{subsec:building}
\begin{center}
\begin{figure}[t]
\includegraphics[width=\linewidth]{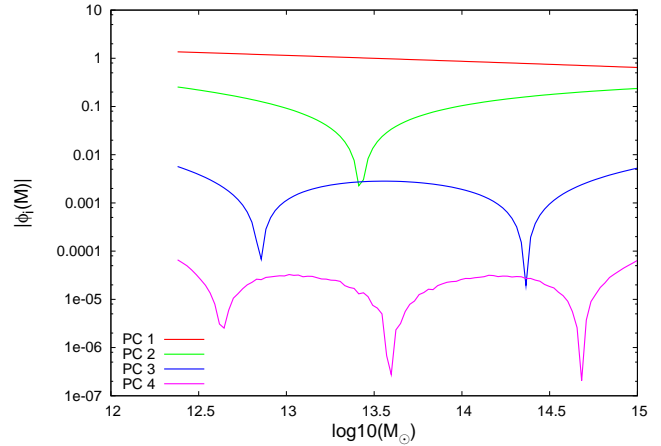}
\caption{First four PC basis vectors, $\phi_i$. Absolute values are
  used to show the dynamic range on a logarithmic scale. Only the
  first three basis vectors are actually used in the emulation; any
  others contribute on a scale many orders of magnitude smaller.}
\label{fig:PC}
\end{figure}
\end{center}

In this section, we briefly outline the process for building the $c-M$
emulator. We follow the procedure explained in~\cite{coyote2} and
refer the reader to this paper for more complete details. The focus
of~\cite{coyote2} was on modeling the matter power spectrum rather
than the $c-M$ relation, however, the process is essentially
unchanged. Starting with the design of 37 models given in
Table~\ref{tab:basic}, we measure the $c-M$ relation for each
cosmology at $z=0$ and $z=1$ and fit these with a power law as
described in Section~\ref{sec:models} to obtain a smooth functional
form. First, for every mass bin, the global mean value is subtracted,
and then via a simple rescaling, the concentrations are normalized to
have unit variance. This produces a zero-mean, unit variance dataset
spanning the 37 cosmologies. To reduce the dimensionality of the
problem, these normalized functions are then decomposed into principal
component (PC) basis functions and only the most significant
components are kept. The idea is to apply the interpolation method of
choice (Gaussian process modeling in our case) to the coefficients of
the basis functions, rather than to the raw data itself
(see~\citealt{coyote2} for details). Figure~\ref{fig:PC} shows that we
only need three PCs to successfully capture the behavior of the $c-M$
relation since the shape of the relationship remains fairly simple
across this set of cosmologies. As explained above, the emulator
actually returns the weight on each PC basis function and these can be
combined together to give the new $c-M$ relation. We model the error
in the projection to the PC basis with an additional hyperparameter
$\lambda_p$ that can be tuned to represent the level of noise in the
data.

A Gaussian process is then used to interpolate between the model
results; this means that the $c-M$ relation for a new cosmology is
actually a function drawn from a unit normal distribution. The
covariance matrix describes the `distance' between the new model and
the set of known models as given by the covariance function. The full
covariance matrix, $\Sigma$, is composed of one $\Sigma_l$ for each
PC, arranged along the diagonal elements such that: $\Sigma =
diag(\Sigma_{1}...\Sigma_{n})$ for $n$ PCs.  Each element of
$\Sigma_l$ is given by:
\begin{equation}
\Sigma_{l;ij} = \lambda_l\prod_{k=1}^{5} \rho_{kl}^{4(\theta_{ik}-\theta_{jk})^2}, 
\end{equation}
where $\theta_l$ represents the cosmological parameters and the $i$
and $j$ indices run over the number of models spanning the design
space (in this case $i,j=1-37$), the $l$ index runs over the number of
PCs and the $k$ index runs over the number of cosmological
parameters. The hyperparameters, $\lambda_l, \rho_{kl}, \lambda_p$,
are set by exploring the likelihood surface, which is done with a
Markov chain Monte Carlo analysis, but any other algorithm that
locates the maximum likelihood of a multidimensional surface could
also be used. The complete expression for the posterior can be found
in Equation B17 of~\cite{coyote2}. This conditions the Gaussian
process to the design of the 37 models and ensures that the
hyperparameters correctly capture the complexity of the surface,
because they control the fit of the interpolating functions to the
data.

After conditioning the GP for the best-fitting hyperparameters, the
emulator is ready to predict the $c-M$ relation for a different
cosmology. The prediction involves re-calculating the covariance
matrix between the new parameters and the design and this locates the
new parameters within the design space. This process is quite fast,
and can be repeated each time a new cosmology is needed with little
computational cost.

The results at intermediate redshifts ($0 < z < 1$) are produced with
a simple linear interpolation. This remains fairly accurate because
the change in the $c-M$ relation with redshift is largely a simple
shift in amplitude.

\subsection{Testing the Emulator}\label{subsec:testing}
The accuracy of the emulator is determined using two methods: 1) we
compare the performance of the emulator against a model not included
in the original design and 2) we remove one of the models from the
design and rebuild the emulator based on the remaining 36 models in
what is known as a holdout test. In this section, we perform both of
these tests to demonstrate the accuracy of the $c-M$ emulator.

\begin{figure}[t]
\includegraphics[width=1\linewidth]{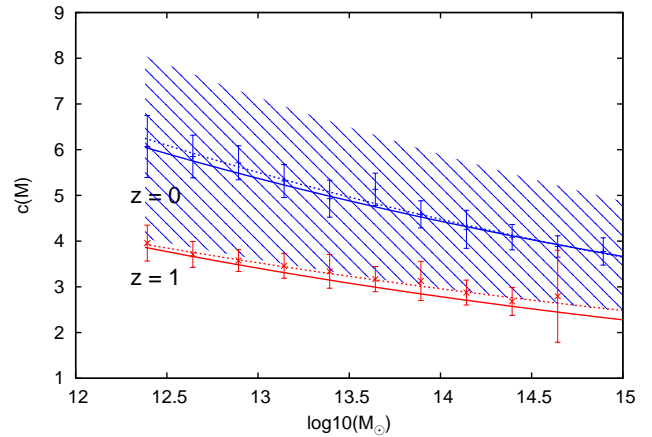}
\caption{Predictions from the $c-M$ emulator at $z=0$ (blue, solid)
  and $z=1$ (red, solid) for measurements from $N$-body simulations
  for the M000 cosmology. The dashed lines show the best-fit power law
  describing the $N$-body results and the hashed region shows the
  expected variation in the $c-M$ relation from the mean (solid
  line). Note that this set of simulations was not used to build the
  emulator.} 
\label{fig:M000}
\end{figure}

We withheld one model (M000) with a $\Lambda$CDM concordance cosmology 
from the set of 37 models when building the $c-M$
emulator. Figure~\ref{fig:M000} shows the comparison between the
emulator prediction for this cosmology against the direct simulation
results from three different box sizes at $z=0$ and $z=1$. The hashed
region covers the 1-sigma boundary around the mean. The emulator
predictions are consistent with the $N$-body $c-M$ relations well
within the errors on the measurements. In comparison with the smoothed
fit for model M000, derived from the same power law fitting procedure
used on the set of 37 cosmologies, we find that at $z=0$, the emulator
is essentially perfect at the high mass end and accurate to at least
3.25\% for low mass halos. For $z=1$ the error is somewhat worse,
mainly due to the limited halo statistics for building the emulator,
especially for the low-$\sigma_8$ models. At low masses the
predictions are accurate at the 2\% level and degrade to 9\%
inaccuracy at the highest masses considered. All of these values are
well within what may be considered to be the nominal uncertainty in
determining concentrations from simulations~\citep{bhattacharya11_b}.
For most of the range of halo mass considered, the accuracy of the
emulator outperforms any other prediction scheme available, especially
considering the large model space covered here.

\begin{figure}[t]
\includegraphics[width=\linewidth]{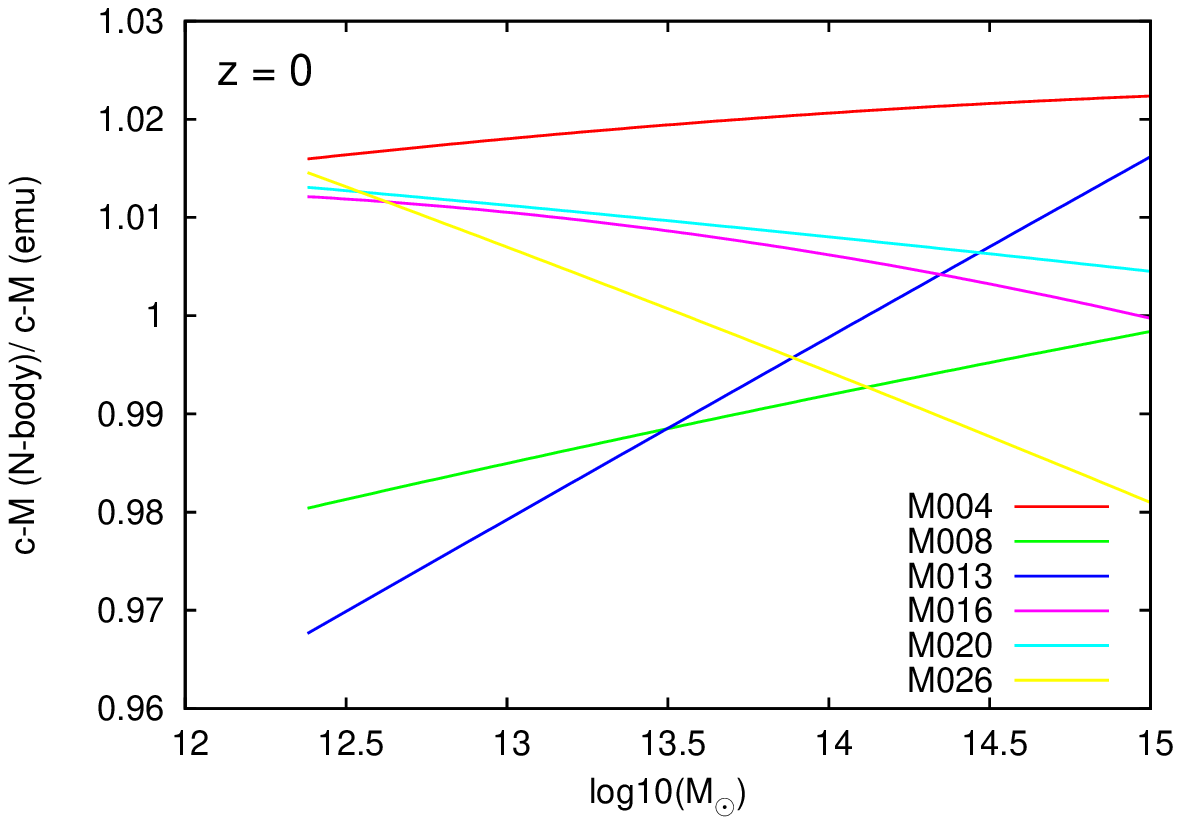}
\includegraphics[width=\linewidth]{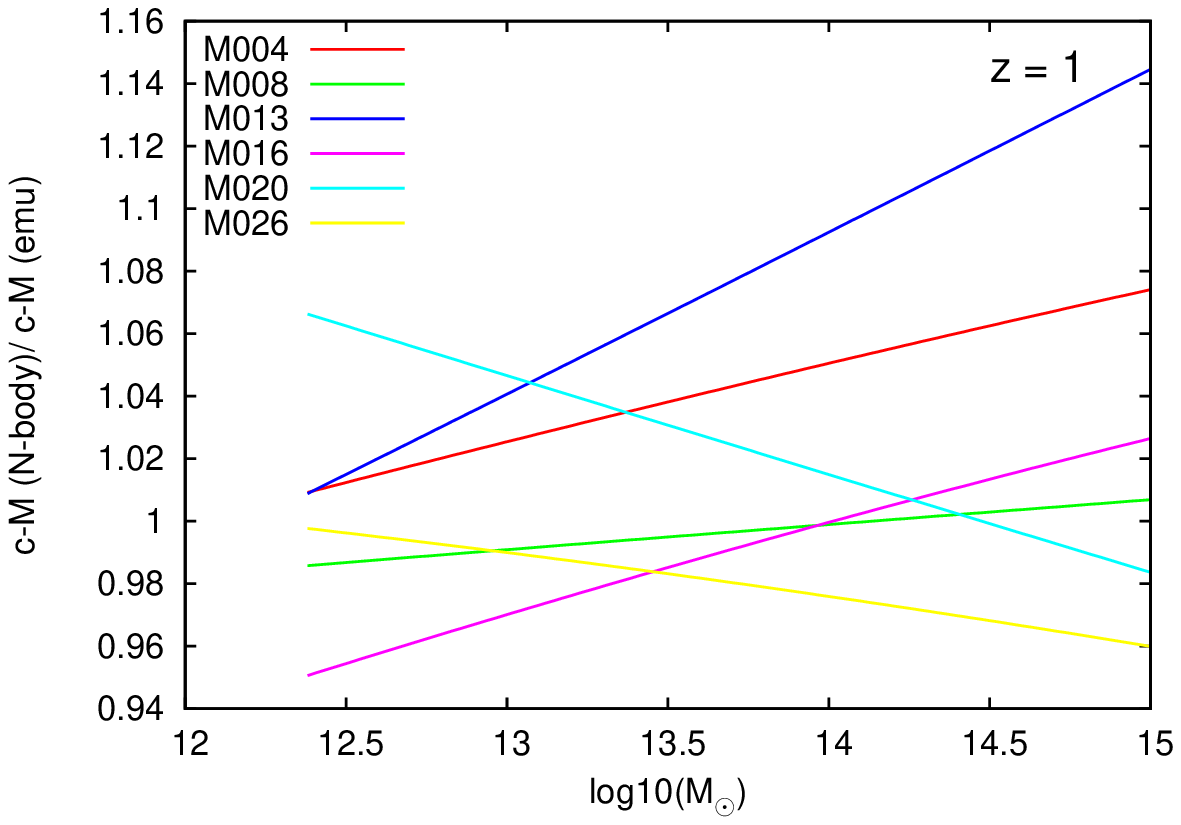}
\caption{Holdout tests for the $c-M$ emulator at $z=0$ (top) and $z=1$
  (bottom). In both plots, the labelled model has been removed from the
  design and an emulator is rebuilt on the reduced design. We then
  take the ratio of the smoothed $N$-body result and the prediction from
  the new emulator to check the accuracy of the full emulator made
  with the original design.}
\label{fig:holdouts}
\end{figure}

In Fig.~\ref{fig:holdouts}, we show estimates of the emulator error by
performing a holdout test. In such a test one model is kept aside and
a new emulator is built, based on the remaining 36 models.  The new
emulator is used to predict the $c-M$ relation for the held-out
model. Since the numerical result (`truth') is known for that model,
we can measure the emulator prediction error. One shortcoming of this
method -- in particular if only a very small number of simulations is
available as is the case here -- is that by removing one model, the
quality of the emulator is degraded.  Therefore, the error estimate
for the emulator obtained this way can be considered to be a
conservative upper bound.

We have chosen to exclude only models M004, M008, M013, M016, M020 and
M026, because these are located relatively close to the center of the
design. Removing a model that defines one of the edges of the design
would greatly reduce the performance of the emulator, since the GP
would be extrapolating for a missing model that is now outside of the
design range. The comparison is made with respect to the smoothed
$N$-body result that was used to construct the full emulator, not the
raw concentration measurements from the simulation. At most, the
emulator deviates by 3.3\% from the simulation results at $z=0$ and
this rises to 15\% at $z=1$. This is because the error on the raw
measurements increases with redshift as the sample size of halos
decreases, particularly for low-$\sigma_8$ models.

\subsection{Comparison with other $c-M$ Predictions}

\begin{figure}[t]
\includegraphics[width=\linewidth]{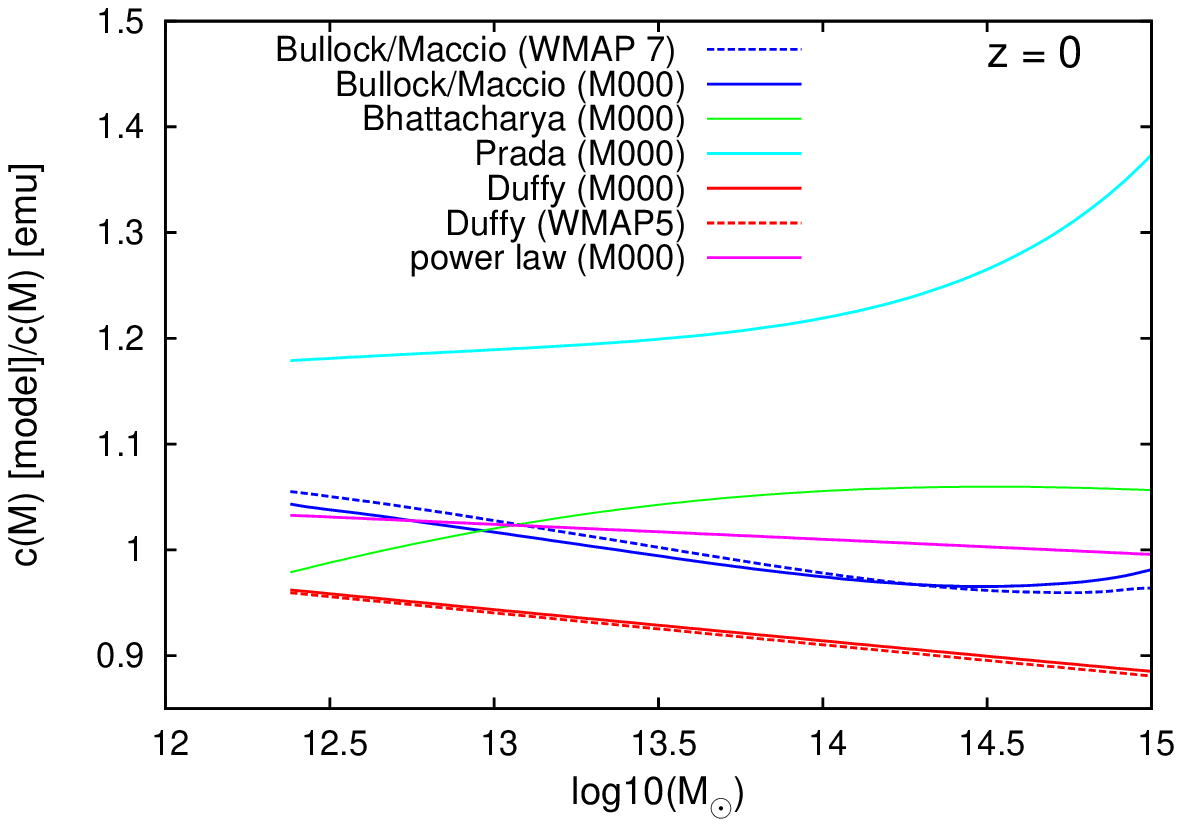}
\includegraphics[width=\linewidth]{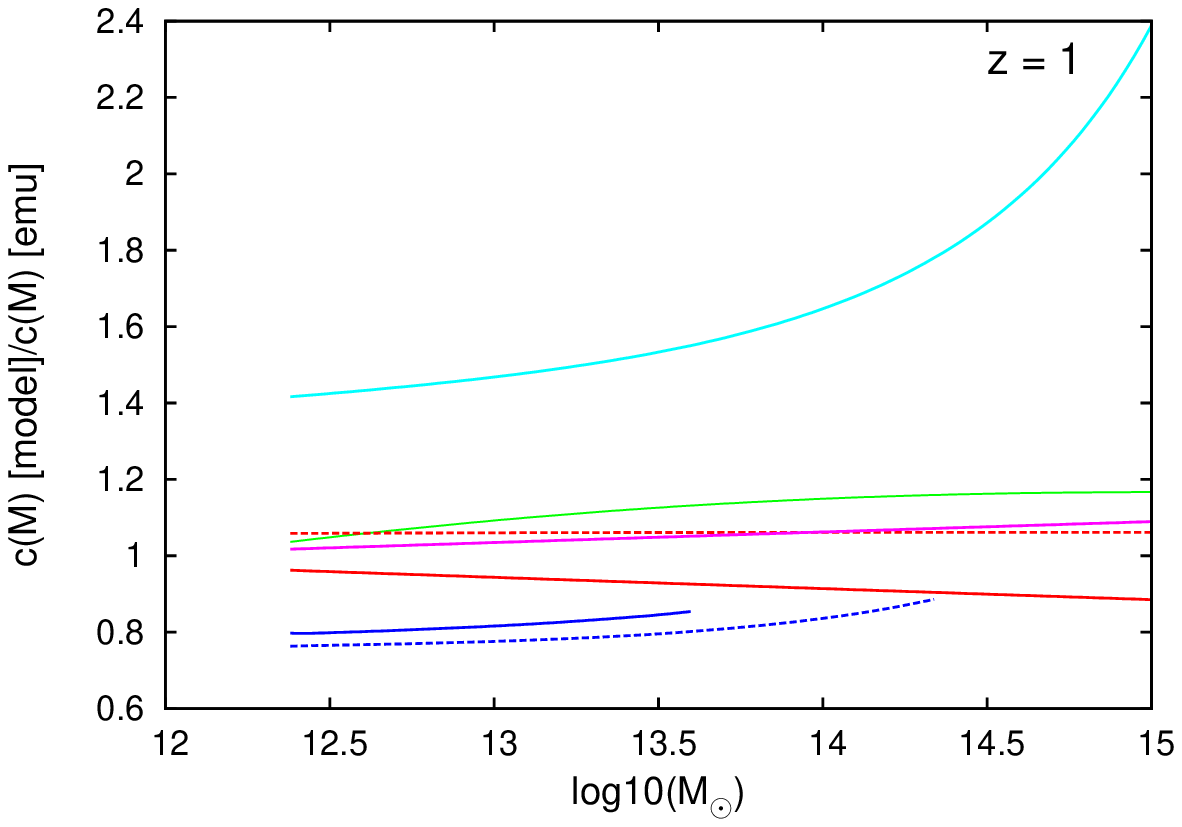}
\caption{Comparison of the emulator -- taken as the reference -- with
  other models for the $c-M$ relation at $z=0$ (upper panel) and $z=1$
  (lower panel). The \cite{bullock}/\cite{maccio08} $c-M$ relation is
  shown in blue for both M000 (solid) and WMAP7 (dashed)
  cosmologies. The \cite{bhattacharya11_b} fit (note that this fit was
  derived only for M000, not for general cosmological models) is in
  green, the~\cite{prada12} $c-M$ relation in cyan, and
  the~\cite{duffy08} $c-M$ relation in red for M000 (solid) and WMAP5
  (dashed) cosmologies. In addition, the pink line shows the ratio of
  the power-law fit for M000 to the emulator prediction (As shown in
  Fig.~\ref{fig:M000} the agreement is very good, with less than 3\%
  deviation over most of the mass range). The lower panel shows the
  results for $z=1$. The publicly available code for
  the~\cite{bullock}/\cite{maccio08} fit does not work seamlessly over
  the full mass range so we do not show results for the very high mass
  end here. See the text for further discussion of this set of
  results.}
\label{fig:bullock}

\end{figure}

We now compare the results obtained from the emulator with those from
the models presented in~\cite{bhattacharya11_b},
~\cite{bullock},~\cite{duffy08} and~\cite{prada12}. The~\cite{bullock}
model was intended to correct the redshift dependence of the original
NFW model, which was claimed to overpredict the concentration of high
redshift ($z > 1$) halos. We perform our comparisions against the most
recent version of the model that incorporates corrections from
\cite{maccio08}\footnote{available from {\tt
    physics.uci.edu/$\sim$bullock/CVIR}}. The~\cite{bullock} model
contains two free parameters $K = 3.85$ and $F =0.01$. Newer values of
$K$ and $F$ were obtained in \cite{maccio08} by fitting this model to
$N$-body simulations using cosmological parameters corresponding to
the first, third and fifth WMAP data
releases. Figure~\ref{fig:bullock} shows the ratio of
the~\cite{bullock} model to our emulator at $z=0$ for two cosmologies,
M000 and WMAP7~\citep{wmap7}; note that the~\cite{bullock} model has
only been tested with $\Lambda$CDM and SCDM cosmologies. These two
models are certainly consistent at low halo masses, within the
expected error of the emulator, but a substantial discrepancy occurs
at cluster-sized halos, even with the updated version
of~\cite{maccio08}. This occurs because the model contains free
parameters that need to be tuned to a particular cosmology with
$N$-body simulations. However, the~\cite{bullock} model is able to
reach much lower halo masses, $M < 10^{10}$ M$_\odot$, than our
emulator because it is calibrated to higher mass resolution $N$-body
simulations.

At $z=1$, the public code used for the Bullock/Macci\'o model fails to
compute the concentration across the full range of halo masses because
of difficulties at low $\sigma_8$. We therefore show only results for
a limited mass range. The discrepancy here is much larger than for
$z=0$, with a concentration underestimation of greater than 20\%.

More recently,~\cite{duffy08} proposed a new $c-M$ relation with a
power-law relationship between the halo mass and the concentration, as
extracted from a series of high resolution, small to medium volume
$N$-body simulations with a WMAP5 cosmology. Results from the $c-M$
emulator are consistent with their predictions to within $\sim$ 10\%
at $z=0$, as are the results in~\cite{bhattacharya11_b}. There is a
slight deviation at cluster sized halos; our $c-M$ emulator is based
on larger volume simulations, and is therefore able to provide a more
complete sample of massive halos and reduced shot noise at the high
mass end. The agreement between the emulator and the~\cite{duffy08}
$c-M$ relation improves to $\sim$ 5-6\% at $z=1$ for the WMAP5
cosmology.

\begin{center}
\begin{figure*}
\includegraphics[width=1.75in,angle=270]{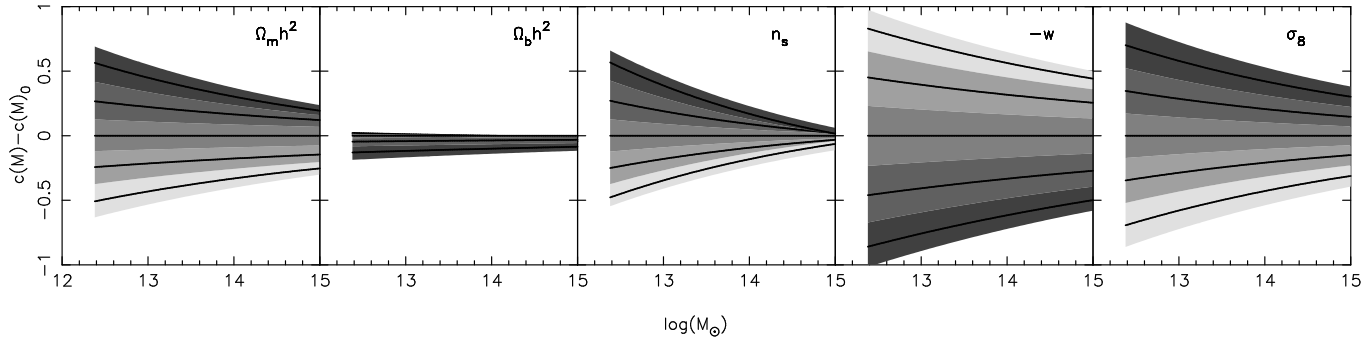}
\caption{Sensitivity of the $c-M$ relation to the five cosmological
  parameters varied in the emulator design, $\Omega_mh^2,
  \Omega_bh^2,-w, n_s$ and $\sigma_8$, at $z=0$. We vary each
  cosmological parameter individually for each panel and have binned
  the range into five intervals, which are coloured from light to dark
  as the value of the parameter increases. From each of these, we
  subtract the $c-M$ relation for the model corresponding to the
  midpoint of the design space, c(M)$_0$. To guide the eye, we also
  plot the median c(M) - c(M)$_0$ of each bin in parameter space.}
  \label{fig:sensitivity}
\end{figure*}
\end{center}

In Figure~\ref{fig:bullock}, we also show the ratio between our
emulator and the $c-M$ relation as determined by the model discussed
in~\cite{prada12}, which is itself based on a number of $N$-body
simulations. There is a $\sim$ 20\% discrepancy at $z=0$ ($\sim$ 40\%
at $z=1$) for lower halo masses. This increases dramatically for
cluster sized haloes at both redshifts, since unlike~\cite{prada12},
we do not observe an upturn in the $c-M$ relation, where their
concentration increases with halo mass. One should note that the
methods for measuring the halo concentration are different in our two
cases -- we use a finite-range profile-fitting method as discussed
in~\cite{bhattacharya11_b}, whereas~\cite{prada12} use a two-point
ratio method. Exectations for discrepancies between profile fitting
and their particular ratio method are further discussed in the
appendix of~\cite{bhattacharya11_b}.

We also note that there is a good agreement to within $\sim$5\%
between our emulator and the $c-M$ relations measured by~\cite{neto07}
from the Millennium simulation~\cite{springel05}. The cosmology of the
Millennium simulation does not quite fall within the range of our
emulator ($\omega_bh^2 = 0.024$ and $h = 0.73$), and to facilitate
this comparsion, we have adoped values as close to these as possible
that still lie within our parameter space ($\omega_bh^2 = 0.0235$ and
$h = 0.719$).

Lastly, we compare the emulator prediction with the fitting function
derived in \cite{bhattacharya11_b} for the M000 cosmology. Before
doing so, we provide some necessary background. First, the redshift
dependence in the fitting form in \cite{bhattacharya11_b} is handled
differently than in the current paper. In \cite{bhattacharya11_b}, the
aim was to find a global power-law fit that encompasses all redshifts
considered (between $z=0$ and $z=2$) at once.  Therefore, the fit for
each redshift is not expected to be perfect. In the current paper we
follow a different path: since we do not provide a single formula for
the $c-M$ relation but rather a simple numerical code, we can generate
the best-fit power-law model for each redshift separately and then
simply interpolate between the redshifts. This produces a more
accurate answer at each redshift at the minimal cost of running a fast
code for every $c-M$ prediction instead of using one fitting formula.

Second, for the high-mass range, \cite{bhattacharya11_b} used higher
force resolution simulations. As shown in the Appendix of
\cite{bhattacharya11_b}, the concentrations from the Coyote runs are
slightly lower at high masses (at the 5\% level) compared to
higher-resolution simulations. Since it is not clear if this effect is
independent of cosmology (most likely for lower $\sigma_8$ simulations
the effect will be smaller) we decided to not attempt to correct the
concentration measures in this paper for the Coyote runs. Therefore,
the uncertainty for the high mass concentrations from the emulator
predictions will be slightly higher and one expects the predictions to
be biased slightly low. Considering the overall scatter and
uncertainty in the $c-M$ relation, this small effect is unlikely to be
significant. One should note, however, that due to this suppression,
the ratio of $\sigma_c(M)$ to the mean concentration, which we quote
at the nominal value of 1/3, could be slightly smaller at higher
masses.

Keeping these caveats in mind, we now turn to the comparison of the
fit by \cite{bhattacharya11_b} and the emulator result, in
Fig.~\ref{fig:bullock}. For $z=0$, both agree at the 2\% (low halo
mass) to 6\% (high halo mass) level, the \cite{bhattacharya11_b} fit
being slightly higher as expected.  For $z=1$, the discrepancy ranges
from 4\% to 17\%. Here, overall the emulator estimate compared to the
best-fit power-law to the simulation result is slightly low, while the
\cite{bhattacharya11_b} fit slightly overestimates the simulation
results. In other words, the actual simulation result lies in between
the emulator prediction and the fit. Overall, the agreement between
the \cite{bhattacharya11_b} fit and the emulator is much better at
$z=1$ than the agreement between the emulator and the Bullock/Macci\'o
fit.

\subsection{Cosmology Dependence of the $c-M$ Relation}

Finally, we explore the sensitivity of the $c-M$ relation to
variations in cosmology. Since we now have a means of quickly and
smoothly interpolating from one cosmology to another, we can simply
vary the parameters that we incorporated into our design space in a
regular grid. We divide each parameter range into five evenly spaced
regions and vary only a single parameter at a time by keeping the
other four parameters fixed at the midpoint of the parameter
range. Our comparisons are always made with respect to this model at
the midpoint, which has the parameters: $\Omega_mh^2 = 0.1375$,
$\Omega_bh^2 = 0.0225$, $n_s = 0.95$, $w = -1$ and $\sigma_8 = 0.758$,
and is subtracted from each $c-M$
relation. Figure~\ref{fig:sensitivity} shows the results of this
exercise at $z=0$, with the values of the cosmological parameters
increasing as the shading increases from light to dark. The entire
region is coloured to show the variation in the $c-M$ relation across
each parameter bin. The range of concentration variation changes as a
function of mass and the cosmological parameter being varied; the
largest variation is of order unity. Unsurprisingly, the $c-M$
relation increases with $\Omega_mh^2$ and $\sigma_8$ and decreases
with $w$, which slows the rate of structure formation. We see little
variation with $\Omega_bh^2$ because the range of parameters allowed
by the CMB constraints is already quite tight. Also, cluster-sized
halos appear to be relatively less sensitive to these changes in
cosmology. Figure~\ref{fig:sensitivity} also shows a clear degeneracy
between $\Omega_mh^2$, $\sigma_8$, and $n_s$ in the $c-M$ relation.

\section{Conclusion and Outlook}
\label{sec:conc}

In this paper, we have presented a new prediction scheme -- in the
form of an emulator -- for the $c-M$ relation for dark
matter-dominated halos at the bright galaxy to cluster mass scales,
covering a range of $2\cdot 10^{12}$M$_\odot<M<10^{15}$M$_\odot$ and a
redshift range of $z=0$ to $z=1$. The emulator provides results for a
large class of $w$CDM cosmologies and is accurate at the $\sim5\%$
level (better for lower redshifts, slightly worse for higher
redshifts). The emulator enables consistent predictions to be made
when testing for deviations from $\Lambda$CDM using clusters. This is
particularly important for cluster cosmology, since the behaviour of
the $c-M$ relation can vary by as much as 30\% just by varying the
equation of state across the range $-1.3 < w<-0.7$. By correctly
including the cosmology dependence in the $c-M$ relation, the emulator
improves on analytic modelling of halo profiles, such as the 1-halo
term used in the halo power spectrum. The performance of the emulator
compares favourably with the other models for the $c-M$ relation in
the literature and outperforms the Bullock/Macci\'o model across the
redshift and mass range considered.

Aside from predicting the mean $c-M$ relation, the interesting and
useful fact that across all 37 cosmologies considered, 1) the scatter
in halo concentrations in individual mass bins is Gaussian, and 2) the
corresponding standard deviation is given by roughly a third of the
mean concentration value, means that the $c-M$ emulator also includes
within it the information regarding the concentration distribution at
a given value of mass.

The work in this paper is an example of how the cosmic calibration
framework provides a means of estimating highly nonlinear quantities
involving evolved structures from a limited number of computationally
expensive $N$-body simulations. In the future, we will extend the
number and range of cosmological parameters to include more exotic
phenomena, such as evolving dark energy, as a complement to upcoming
dark energy experiments.

\begin{acknowledgments}

  The work at Argonne National Laboratory was supported under
  U.S. Department of Energy contract DE-AC02-06CH11357.  Part of this
  research was supported by the DOE under contract W-7405-ENG-36. We
  are indebted to Charlie Nakhleh for providing an example code for
  building cosmic emulators and Earl Lawrence and Dave Higdon for many
  useful and entertaining discussions on the topic. We thank Volker
  Springel for making {\sc GADGET-2} publicly available.  We are
  grateful for computing time granted to us as part of the Los Alamos
  Open Supercomputing Initiative. This research used resources of
  the Argonne Leadership Computing Facility at Argonne National
  Laboratory and the National Energy Research Scientific Computing
  Center, which are supported by the Office of Science of the
  U.S. Department of Energy under contract DE-AC02-06CH11357 and
  DE-AC02-05CH11231 respectively.

\end{acknowledgments}

\end{document}